\documentclass[aps,prd,amsmath,showpacs,preprintnumbers,superscriptaddress,
               twocolumn]{revtex4}
\usepackage{amsmath}
\usepackage{graphicx}
\usepackage{latexsym}

\newcommand{\beqa}{\begin{eqnarray}}
\newcommand{\eeqa}{\end{eqnarray}}
\newcommand{\beqn}{\begin{equation}}
\newcommand{\eeqn}{\end{equation}}

\def\spose#1{\hbox to 0pt{#1\hss}}
\def\ltapprox{\mathrel{\spose{\lower 3pt\hbox{$\mathchar"218$}}
 \raise 2.0pt\hbox{$\mathchar"13C$}}}
\def\gtapprox{\mathrel{\spose{\lower 3pt\hbox{$\mathchar"218$}}
 \raise 2.0pt\hbox{$\mathchar"13E$}}}

\begin{document}

\title{Positivity violation for the lattice Landau gluon propagator}
\author{Attilio Cucchieri}
\author{Tereza Mendes}
\affiliation{Instituto de F\'\i sica de S\~ao Carlos,
Universidade de S\~ao Paulo, \\
C.P.\ 369, 13560-970, S\~ao Carlos, SP, Brazil}
\author{Andre R.\ Taurines}
\affiliation{Instituto de F\'\i sica Te\'orica,
Universidade Estadual Paulista,\\
Rua Pamplona, 145, S\~ao Paulo, SP, Brazil
}

\begin{abstract} 
\noindent
We present explicit numerical evidence of reflection-positivity
violation for the lattice Landau gluon propagator in three-dimensional 
pure $SU(2)$ gauge theory. 
We use data obtained at very large lattice 
volumes ($V = 80^3, \, 140^3$) and for three different lattice
couplings in the scaling region ($\beta = 4.2, \, 5.0, \, 6.0$).
In particular, we observe a clear oscillatory pattern in the
real-space propagator $C(t)$. 
We also verify that the (real-space) data show good scaling in the 
range $t \in [0, 3]\,fm$ and can be fitted using a Gribov-like form.
The violation of positivity is in contradiction with
a stable-particle interpretation of the associated field theory
and may be viewed as a manifestation of confinement.
\end{abstract}


\maketitle

\section{Introduction}

In recent years, there has been considerable interest in the
possible violation of spectral positivity for QCD and in its 
relation to color confinement (for a review see \cite{Alkofer:2000wg}).
Let us recall \cite{Haag:96} that the reconstruction of a 
G\aa rding-Wightman relativistic quantum field theory from the
corresponding Euclidean Green functions is possible only if they
obey the Osterwalder-Schrader axioms \cite{axioms}. 
In particular, the requirement
of positive definiteness of the norm in Hilbert space
is expressed in Euclidean space by the axiom
of {\em reflection positivity}. For a generic 2-point function
${\widetilde D}(x-y)$, this axiom reads
\begin{equation}
\int d^4x \, d^4y \, f^*(-x_0,{\mathbf{x}}) \,
{\widetilde D}(x-y)\,f(y_0,{\mathbf{y}})\;\ge\; 0~,
\label{eq:violationcontinuum}
\end{equation}
where $f(x_0,{\mathbf{x}})$ is an arbitrary complex test function.
The above condition implies the existence of a K\"allen-Lehmann
representation for ${\widetilde D}(x-y)$, which is necessary for interpreting 
the fields in terms of stable particles. 
Thus, a violation of (\ref{eq:violationcontinuum}) implies that
the Euclidean 2-point function cannot represent the correlations of
a physical particle. This may be viewed as an indication of
{\em confinement} \cite{Alkofer:2000wg}.

The relation between reflection positivity and Euclidean 
correlation functions can be made explicit by considering
the spectral representation \cite{Alkofer:2000wg,Aiso:au}
\begin{equation}
D(p)\;=\; \int_0^{\infty}\,dm^2\,\frac{\rho(m^2)}{p^2 + m^2}
\end{equation}
for the Euclidean propagator in momentum space.
Then, the statement of reflection positivity is equivalent to a 
positive spectral density $\rho(m^2)$.
This implies that the temporal correlator at zero spatial momentum
$D(t, {\bf p}=0)$ can be written as
\begin{equation}
C(t)\;\equiv\;D(t,0)\;=\; 
\int_0^{\infty}\,d\omega\,\rho(\omega^2)\,e^{-\omega\,t}\;.
\label{eq:Cdecomposition}
\end{equation}
We note that for general spatial momentum ${\bf p}$ one would have
$\omega=\sqrt{{\bf p}^2+m^2}$. [In the particular case ${\bf p}=0$ 
considered here, the decay behavior of $\,D(t, {\bf p})\,$ provides
direct insight on mass-like properties associated with the fields.]
Clearly, a positive density $\rho(\omega^2)$ implies that
\begin{equation}
C(t)\;>\;0\;.
\label{eq:condition}
\end{equation}
Notice that having $C(t)>0$ for all $t$ does not ensure the positivity
of $\rho(\omega^2)$. On the other hand,
finding $C(t)<0$ for some $t$ implies that
$\rho(\omega^2)$ cannot be positive, suggesting 
confinement for the corresponding particle.

For the gluon, the Landau propagator is 
predicted to vanish at zero momentum 
\cite{Zwanziger:gz,Gribov:1977wm,Zwanziger:1990by,Zwanziger:1993dh,IRkappa,
Alkofer:2003jj}.
This implies that the real-space propagator ${\widetilde D}(x-y)$
is positive and
negative in equal measure, i.e.\ reflection positivity is maximally
violated \cite{Zwanziger:gz,Zwanziger:1990by,Zwanziger:1993dh}.
An infrared suppressed Landau gluon propagator has been obtained
in several studies in momentum space \cite{infrared,finiteTSU2,Cucchieri:2001tw,
Cucchieri:2003di}.
Numerical indications of a negative real-space lattice Landau gluon
propagator have been presented in the $3d$ $SU(2)$ case
\cite{finiteTSU2},
in the magnetic sector of the $4d$ $SU(2)$ case at finite temperature
\cite{Cucchieri:2001tw} and, recently, in the $4d$ $SU(3)$ case
for one ``exceptional'' configuration \cite{Furui:2004bq}.
In this work we verify this feature in detail (see Section
\ref{sec:violation}), using data obtained
at very large lattice sizes for the $SU(2)$ case in three
space-time dimensions \cite{Cucchieri:2003di}.
At the same time, we try to fit the numerical data in real
space (see Section \ref{sec:fits})
by considering a sum of Gribov-like propagators
\cite{Zwanziger:gz,Gribov:1977wm}.
Let us recall that an excellent fit of the (momentum-space)
propagator by a Gribov-like formula has been obtained
for the equal-time three-dimensional transverse gluon propagator
in $4d$ $SU(2)$ Coulomb gauge
\cite{coulomb} and for the $3d$ $SU(2)$ Landau
case \cite{Cucchieri:2003di}, while in Ref.\ \cite{Aiso:au}
the (real-space) transverse propagator has been fitted using a
a Stingl-like formula in the $4d$ $SU(3)$ Landau case.
Also, several fitting forms have been considered
in Ref.\ \cite{Leinweber:1998uu} for the gluon propagator in momentum
space.

%%%%%%%%%%%%%%%%%%%%%%%%%%%%%%%%%%%%%%%%%%%%%%%%%%%%%%%%%%%%%%%%%%%%%

\section{Violation of reflection positivity}
\label{sec:violation}

An explicit (numerical) proof of violation of
the condition (\ref{eq:condition}) may be difficult if the
correlation function $C(t)$ only becomes negative at relatively large values
of $t$. In particular, this is the case if
$C(t)$ is of the form $\,C(t) = e^{-\lambda t}\,f(t)\,$ and $f(t)$ is only negative for
$t \gg 1/\lambda$. 
In this case it is helpful to consider alternative quantities $G(t)$ that
are positive if $\rho(\omega^2)$ is positive. 
[Consequently, finding $G(t)<0$ implies violation of positivity for
$\rho(\omega^2)$.]
For example,
one can define \cite{Mandula:nj} the function
\begin{eqnarray}
G(t) &\equiv&
\frac{d^2}{dt^2} \ln C(t)
\label{eq:gofC} \\[2mm]
&=& \frac{C(t) \, C''(t)\,-\, [C'(t)]^2}{[C(t)]^2}  \;.
\label{eq:gofC2}
\end{eqnarray}
Using Eq.\ (\ref{eq:Cdecomposition}) we can write $G(t)$ as
\cite{Alkofer:2000wg,Aiso:au}
\begin{equation}
G(t) \;=\; \langle (\omega \,-\, \langle \omega \rangle )^2 \rangle \;,
\label{eq:goft}
\end{equation}
where the averages denoted by $\langle\;\rangle$
are evaluated in the measure $\,d\omega\,\rho(\omega^2)\,e^{-\omega\,t}$.
Clearly, if the density $\rho(\omega^2)$ is positive, so is $G(t)$.
Let us note that for $C(t) = e^{-\lambda t}\,f(t)$ one gets
$\, G(t) \equiv d^2 \ln f(t) / dt^2 \,$, 
namely we get rid of the exponential factor $e^{-\lambda t}$ and
it should be easy to check numerically if $G(t)$ --- 
or equivalently $f(t)$ --- is negative. The quantity
$G(t)$ could be of particular interest in a $4d$ study,
since in this case it is more difficult
to obtain good data for large time separations.
In the case of a Gribov-like momentum-space
propagator \cite{Gribov:1977wm}
\begin{equation}
D(p) \;=\; p^2\,/\,\left(p^4+M^4\right) \; ,
\label{eq:Gribov}
\end{equation}
one obtains the real-space propagator \cite{Zwanziger:gz} 
\begin{eqnarray}
C(t) &=& \frac{1}{2\,\pi}\,
          \int_{-\infty}^{\infty}\, dp\, D(p) \, e^{-i p t} \\[1mm]
     &=& \frac{e^{-M t / \sqrt{2}}}{2\,M}\,
     \cos\left(\frac{M t}{\sqrt{2}} \,
                      +\,\frac{\pi}{4}\right) \; .
\label{eq:gribovreal}
\end{eqnarray}
Then, using Eq.\ (\ref{eq:gofC2}), it is easy to check that
\beqn
G(t) \; =\; - M^2  \left[2\, \cos^2\left(\frac{M t}{\sqrt{2}} \,
                      +\,\frac{\pi}{4}\right) \right]^{-1} \; ,
\eeqn
which is negative for all values of $t$.

Notice that if $C(t)$ is negative for some
value of $t$ we cannot evaluate its logarithm
in Eq.\ (\ref{eq:gofC}), while the expression in 
Eq.\ (\ref{eq:gofC2}) is always well defined for $C(t)\neq 0$.

\vskip 3mm
On the lattice, the real-space propagator can be evaluated
using
\begin{equation}
C(t) =  \frac{1}{N} \sum_{k_0=0}^{N-1}
e^{- 2\pi\,i k_0 t /N} \,D(k_0, 0) \;,
\label{eq:ltc}
\end{equation}
where $N$ is the number of points per lattice side and $D(k)$ is 
the propagator in momentum space. If the lattice action satisfies
reflection positivity \cite{Montvay:cy},
then we can write the spectral representation
\beqn
C(t) = \sum_n r_n e^{-E_n t} \;,
\label{eq:lattice_coft}
\eeqn
where $r_n$ are positive-definite constants.
Clearly, this implies that $C(t)$ is non-negative for all
values of $t$. 

\begin{figure}[t]
\begin{center}
\includegraphics[height=0.75\hsize]{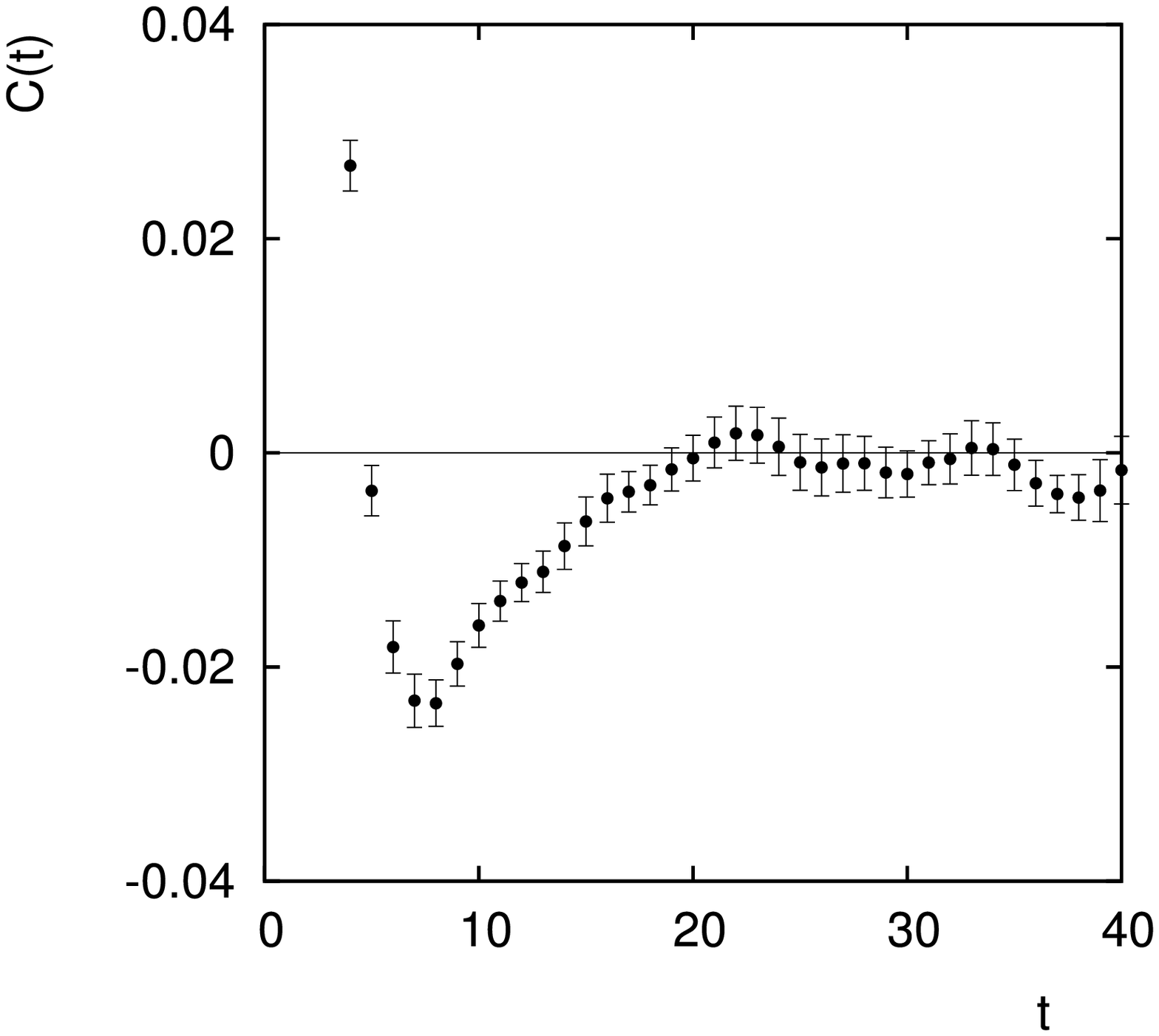}
\protect\hskip -1.2cm
\protect\vskip 0.4cm
\includegraphics[height=0.75\hsize]{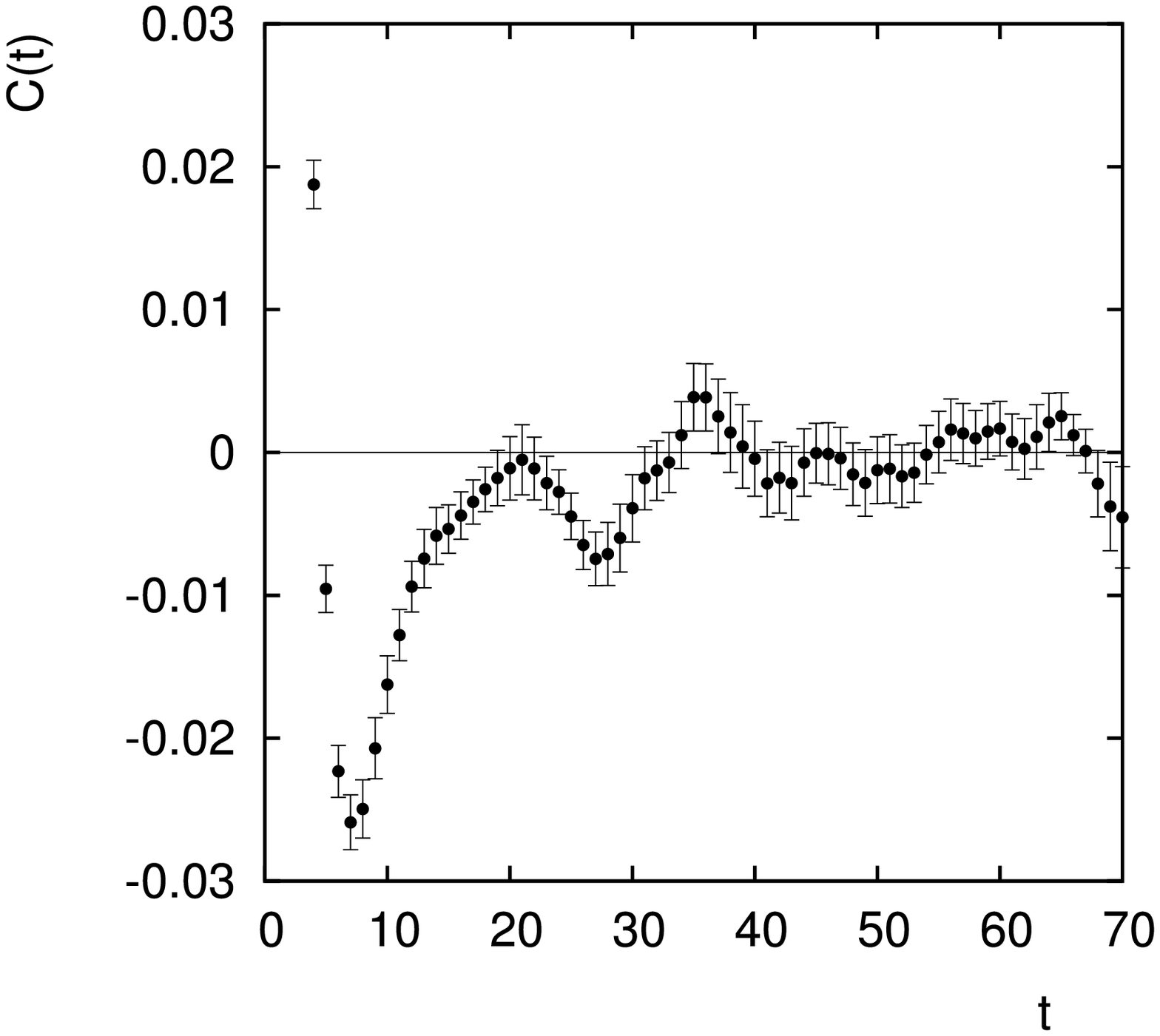}
\protect\vskip -0.4cm
\end{center}
\caption{Real-space propagator $C(t)$ as a function of $t$
for coupling $\beta=5.0$ and lattice volumes $V = 80^3$ (above)
and $V = 140^3$ (below).
Errors have been evaluated using bootstrap with 1000 samples.
All quantities are in lattice units.
\label{fig:redect_zoom}
}
\end{figure}

\begin{figure}[t]
\begin{center}
\includegraphics[height=0.75\hsize]{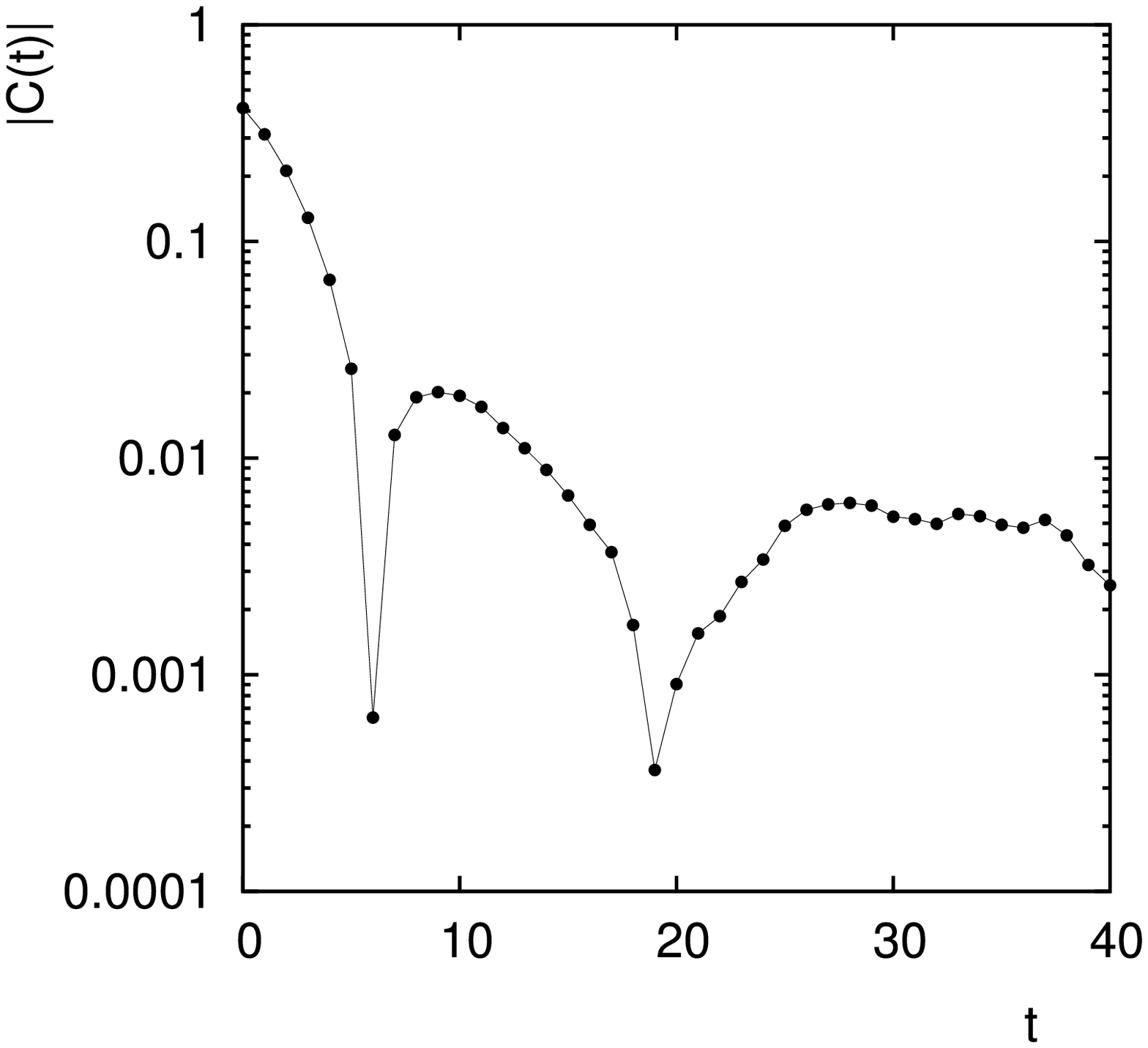} 
\protect\vskip -0.4cm
\end{center}
\caption{Plot of $\,| C(t) |\,$ as a function of $t$
for lattice volume $V = 80^3$ and coupling $\beta=6.0$.
For clarity, errors are not shown.
All quantities are in lattice units.
The solid line is drawn to guide the eye.
\label{fig:redect_zoom_log}
}
\end{figure}
\begin{figure}
\begin{center}
\includegraphics[height=0.75\hsize]{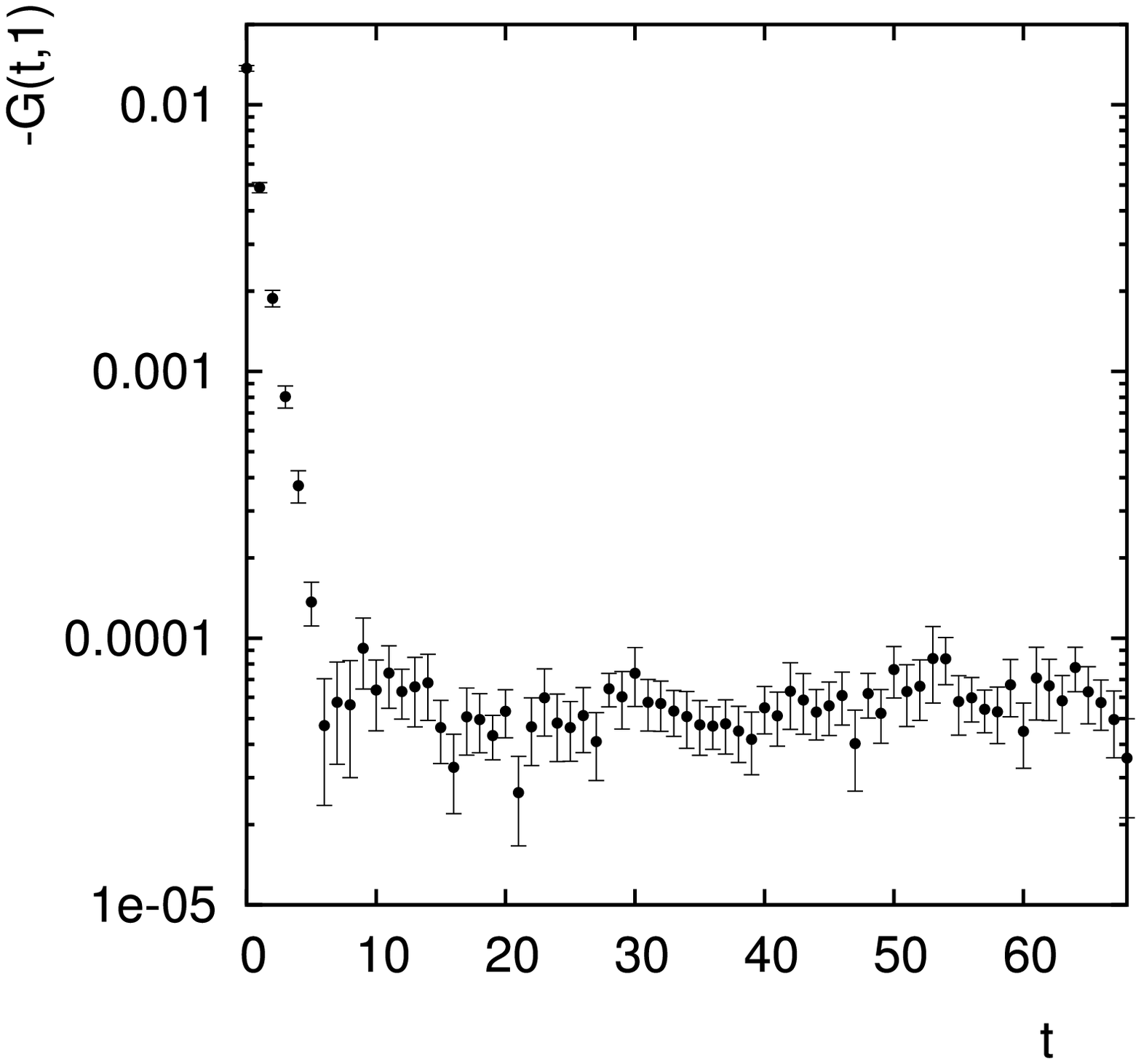}
\protect\vskip -0.4cm
\end{center}
\caption{Plot of $\, - G(t,1) \,$ as a function of $t$
for lattice volume $V = 140^3$ and coupling $\beta=4.2$. Errors
have been evaluated using bootstrap with 1000 samples.
All quantities are in lattice units.
\label{fig:lattice_concavity}}
\end{figure}

As in the continuum, we can consider
$G(t)$ using Eq.\ (\ref{eq:gofC2}) or, equivalently,
the function $G(t) \,[C(t)]^2$. This quantity can be easily
discretized on the lattice by
\beqn
G(t,a) \; =\; 
\frac{1}{a^2}\left[\,C(t)\,C(t+2a) \,-\,
C(t+a)^2 \,\right] \;,
\label{eq:defGlatt}
\eeqn
where $\,a\,$ is the lattice spacing.
Indeed,  in the continuum limit $\,a\to0\,$, one obtains 
$\, G(t,a) = G(t)\,C(t)^2+{\cal O}(a^3)$.
Furthermore, defining
\beqn
b_n \;=\; \sqrt{r_n} \, e^{- E_n t/2}\;, \quad\;
c_n \;=\; \sqrt{r_n} \, e^{- E_n (t+2a)/2} \;,\; \;
\eeqn
we can use the Schwartz inequality
\cite{Mandula:nj} to show that $\, G(t,a) \ge 0 \,$
for all values of $t$ and $a$ if the $r_n$'s are positive.
Also, considering the effective gluon mass 
\beqn
m(t) \;=\; - \log{\left[{C(t+a)}\, /\,{C(t)}\right]} \; ,
\label{eq:m}
\eeqn
one gets $\,e^{m(t)} - e^{m(t+a)} \,=\, G(t,a) \;a^2 \, e^{m(t+a)}\,/\, C(t+a)^2$.
Thus, if $\, G(t,a) \ge 0 \,$ one obtains
$m(t) \ge m(t+a)$, i.e.\ the effective mass should 
decrease when considering a larger time separation $t$.
As discussed in \cite{Mandula:nj}, an increasing effective gluon mass 
has been obtained already in the first numerical studies of the
gluon propagator \cite{first}, suggesting a violation of
reflection positivity. Note, however, that Eq.\ (\ref{eq:m}) is ill-defined
if $\,C(t)\,$ changes sign.

\begin{figure}[t]
\begin{center}
\includegraphics[height=0.75\hsize]{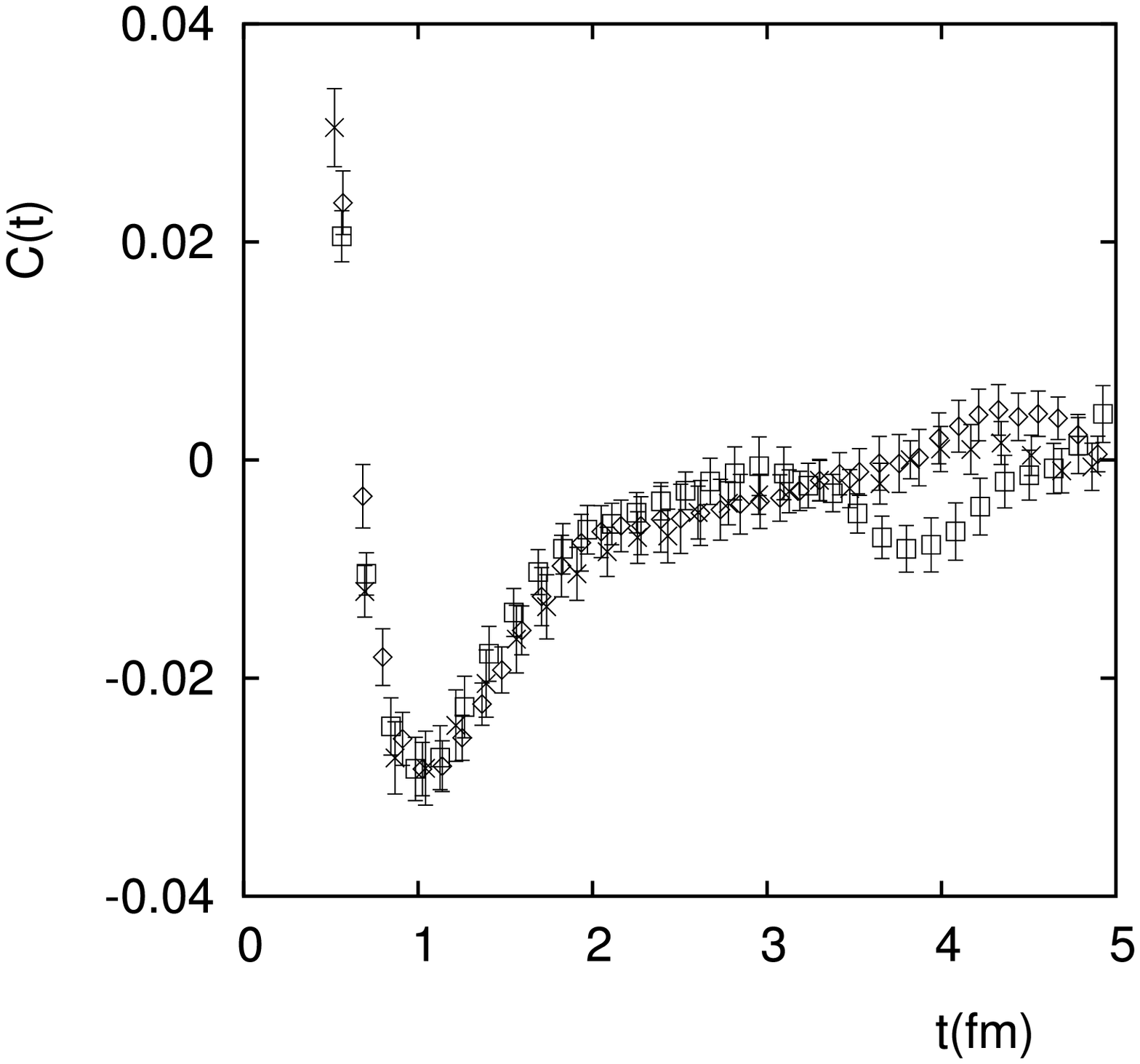}
\protect\vskip -0.4cm
\end{center}
\caption{Scaling for the real-space propagator $C(t)$
as a function of $t$ (in $fm$) for lattice volume $V = 140^3$ and
couplings $\beta=4.2\,(\times)$, $\,5.0\,(\Box)$, $6.0\,(\Diamond)$.
Errors have been evaluated using bootstrap with 1000 samples.
\label{fig:scaling}
}
\end{figure}

Here we use the $3d$ $SU(2)$ Landau-gauge data presented in
\cite{Cucchieri:2003di} in order to check
if the conditions $C(t) >0$ and $G(t,a) > 0$ are
violated for the gluon propagator.
(The data have been analyzed using
the bootstrap method with $1000$ samples; we
checked that results do not change when using $500$ samples.)
As explained in Ref.\ \cite{Cucchieri:2003di}, we set
the physical scale by considering $3d$ $SU(2)$ lattice results
for the string tension and the input value $\sqrt{\sigma} = 0.44 \, 
GeV$, which is a typical value for this quantity in the $4d$
$SU(3)$ case. Since we consider $\hbar = c = 1$, this
implies $1 \,fm^{-1} = 0.4485 \sqrt{\sigma}$.

We find that
the real-space propagator $C(t)$ is negative
for several values of $t$, showing a clear
oscillatory behavior (see Fig.\ \ref{fig:redect_zoom}).
In analogy with Ref.\ \cite{Alkofer:2003jj} we also
plot, in Fig.\ \ref{fig:redect_zoom_log}, the function
$| C(t) |$: the spikes reveal the
change of sign in the propagator $C(t)$.
Finally, in Fig.\ \ref{fig:lattice_concavity} we plot
the function $- G(t,1)$: one can see that, as in
the Gribov-like propagator, $G(t,1)$ is
negative for all values of $t$.
Thus, we find an {\em explicit violation of positivity}
for the lattice Landau gluon propagator. Let us stress
that this violation is clearly observable for the
three lattice couplings and for the two lattice volumes
considered.

%%%%%%%%%%%%%%%%%%%%%%%%%%%%%%%%%%%%%%%%%%%%%%%%%%%%%%%%%%%%%%%%%%%%%

\section{Scaling and Fits for $C(t)$}
\label{sec:fits}

It is important to check if the behavior obtained for $C(t)$
satisfies {\em scaling} for the lattice parameters
considered here. To this end, we apply to the data the matching
procedure described in \cite[Section III]{Cucchieri:2003di} and
consider $t$ in physical units using
\cite[Table 2]{Cucchieri:2003di}. We obtain that all propagators
become negative at $t \approx 0.7\,fm$ and that the minimum
is reached at $t_{min} \approx 1\,fm$ (see Fig.\ \ref{fig:scaling}).
Moreover, finite-size effects seem to become important only
at $t \gtapprox 3\,fm$. This means that our data for $t \in [0, 3]\,fm\,$ 
are essentially infinite-volume continuum results.
Note that the Gribov-like propagator $C(t)$ in Eq.\
(\ref{eq:gribovreal}) has its minimum at
$t_{min} = \pi / (M \,\sqrt{2}\,)$.
Thus, the above result for $t_{min}$ would imply
$M \approx \pi / \sqrt{2}\,fm^{-1} \approx 2.22\,fm^{-1} = 438\,MeV
= 0.995 \sqrt{\sigma} $.
Let us also observe
that the momentum-space Gribov-like propagator $D(p)$
[see Eq.\ (\ref{eq:Gribov})] has its maximum at $p_{max} = M$.
In Ref.\ \cite{Cucchieri:2003di} we obtained $p_{max} = 0.8^{+0.2}_{-0.1}
\sqrt{\sigma} = 350^{+100}_{-50} \,MeV$.

\begin{figure}[t]
\begin{center}
\includegraphics[height=0.75\hsize]{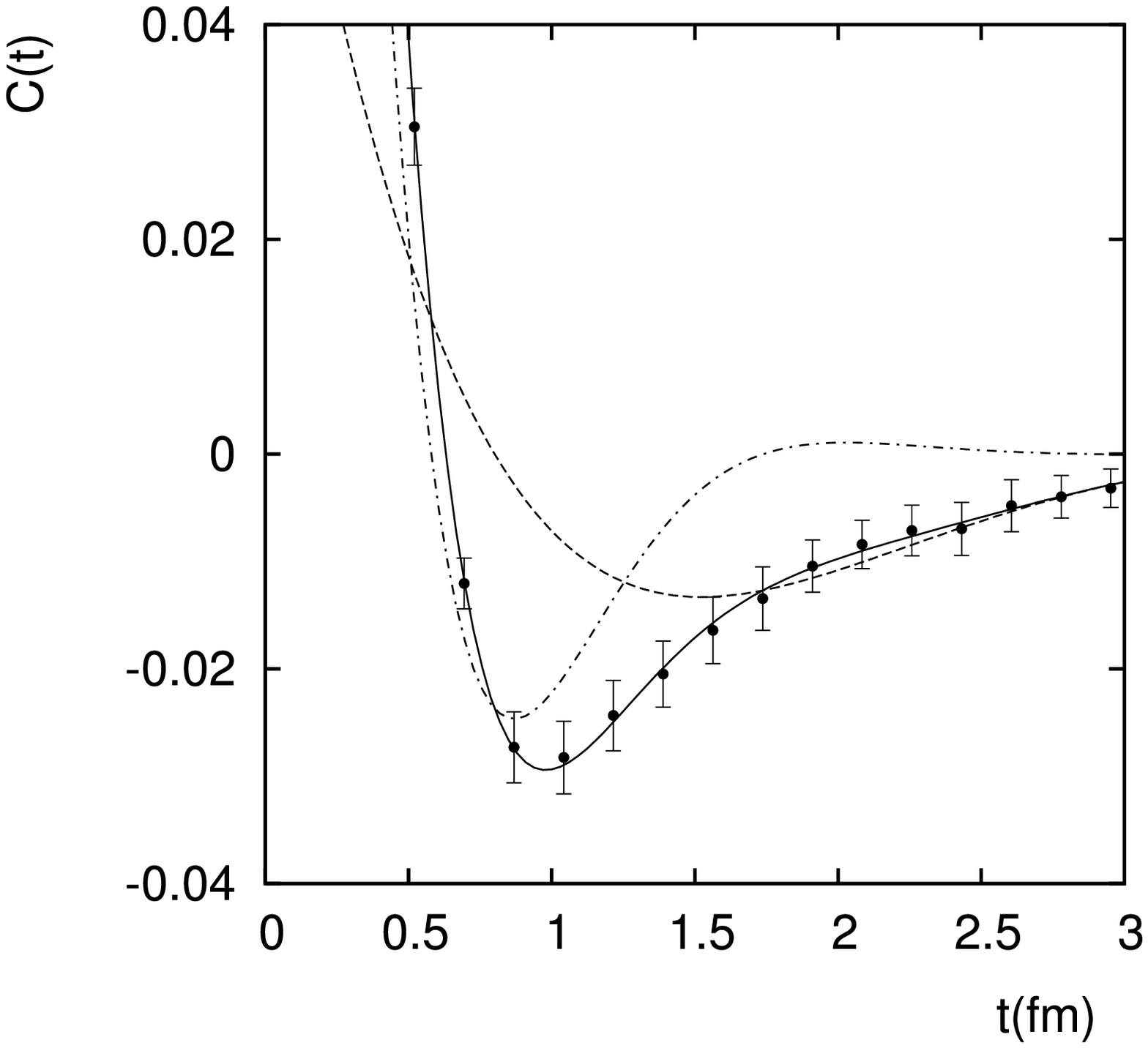}
\protect\vskip -0.4cm
\end{center}
\caption{Fit of ${C}(t)$ as a function of $t$ (in $fm$)
using a sum of two functions of the type (\protect\ref{eq:Cfit})
for lattice volume $V=140^3$ and coupling $\beta=4.2$. We also
display $C_1(t)$ and $C_2(t)$ separately.}
\label{fig:fitting}
\end{figure}

As said above, in Ref.\ \cite{Cucchieri:2003di} we have
fitted the gluon propagator in momentum space using a modified
Gribov-like or Stingl-like formula with four or five parameters.
We now try to fit the data in real space using
\beqn
C(t) \;=\; c\, e^{-\lambda t / \sqrt{2}}\,
     \cos\left(b\, +\,\lambda\, t\,/\, \sqrt{2} \right) \;
\label{eq:Cfit}
\; ,
\eeqn
which is a generalization of the Gribov-like propagator
in Eq.\ (\ref{eq:gribovreal}). Clearly,
this function also corresponds to a $G(t)$ that is
always negative. Let us stress that only for $b = \pi /4$
does this propagator correspond to the Gribov-like propagator
in Eq.\ (\ref{eq:Gribov}). In particular, for $b=0$
one gets the momentum-space propagator
$ D(p) \;=\; \left( p^2 + M^2 \right) / \left( p^4+M^4 \right) \,$,
which is finite at zero momentum.
As reported in \cite{Cucchieri:2003di}, it is still
not clear from our data (on a $140^3$ lattice)
if the zero-momentum gluon propagator vanishes
in the infinite-volume limit, as predicted in
\cite{Zwanziger:gz,Zwanziger:1993dh,Zwanziger:1990by}.

We fit the data obtained for the two largest physical volumes,
i.e.\ $V = 140^3$ and $\beta = 4.2$, 5.0, with $t$ in the range
$[0, 3]\,fm$. As can be seen in Fig.\ \ref{fig:fitting},
the data are well fitted using a sum of two
functions of the type (\ref{eq:Cfit}).
The corresponding fitting parameters are reported in Table
\ref{tab:fit_par}. The averaged mass scales are 
$\lambda_1 = 1.69(1) \sqrt{\sigma} = 745(5)\,MeV$ and $\lambda_2 =
0.74(1) \sqrt{\sigma} = 325(6)\,MeV$.
One can also obtain good fits of the data in the whole $t$
range by considering the Fourier transform of the
sum of three Stingl-like propagators in momentum space.
These fits (using 12 parameters) have been reported
elsewhere \cite{Taurines:2004by}.
It is evident that fits of the gluon propagator in
real space (see also \cite{Aiso:au}) require
more parameters than fits in momentum space.
This is due to the fact that the 
infrared data, for which the modeling
is still not well understood,
are spread over the whole time interval by
the Fourier transform done in the evaluation of
the temporal correlator $C(t)$.

Recently, it has been suggested \cite{Aubin:2003ih,Aubin:2004av} 
that the violation of spectral positivity in lattice Landau gauge be
related to the quenched auxiliary fields used for gauge fixing.
We note that the fitting form proposed for $C(t)$ in \cite{Aubin:2004av}
(also considering 5 fitting parameters)
describes reasonably well our data up to $t=3\,fm$ --- yielding a
light-mass scale of about $1.14 \sqrt{\sigma} = 500\,MeV$ --- but cannot
account for the oscillatory behavior observed at very large
separations.

\begin{table}[th]
\protect\vskip 1mm
\begin{tabular}{c c c c c c}   
$\beta$ & $c_1$ & $\lambda_1$ & $c_2$ & $b_2$ & $\lambda_2$ \\ \hline
4.2 & 0.368(6) & 3.83(4) &  0.70(3) &  0.099(6) &  1.54(4)
    \\ \hline
5.0 & 0.361(6) & 3.72(3) &  0.56(3) &  0.089(6) &  1.75(5)
    \\ \hline
\end{tabular}
\caption{Fit of the data using a sum of two functions of the
type (\protect\ref{eq:Cfit}), setting $b_1=0$.
We obtain $\chi/d.o.f. \approx 0.24$ (respec.\ 0.19) for
$\beta = 4.2$ (respec.\ 5.0).
The number of $d.o.f.$ is 13 (respec.\ 17).
The values of $\lambda_1$ and $\lambda_2$ are in $fm^{-1}$.
The relatively small $\chi/d.o.f.$ is probably due to the use of the
diagonal part of the covariance matrix only.
\label{tab:fit_par}}
\protect\vskip -0.7cm
\end{table}
 
%%%%%%%%%%%%%%%%%%%%%%%%%%%%%%%%%%%%%%%%%%%%%%%%%%%%%%%%%%%%%%%%%%%%%

\section{Conclusions}

Using data from the largest lattice sides to date, we
verify explicitly (in the $3d$ case) the violation of 
reflection positivity for the $SU(2)$ lattice Landau 
gluon propagator. This is one of the manifestations of 
confinement discussed in \cite{Alkofer:2000wg}.
For very large separations ($t> 3\,fm$) the propagator
shows a clear oscillatory behavior, but of course one needs
a careful extrapolation to infinite volume in order to
verify if this behavior survives in that limit.
In the scaling region, the data are well described by 
a sum of Gribov-like formulas, with a light-mass scale 
$M\approx 0.74 \sqrt{\sigma} = 325\,MeV$, where $\sigma$
is the string tension.
As a final comment, one should always bear in mind that the
Gribov-like propagator may not represent the true analytic
structure of the gluon propagator, but it is illustrative of a possible
mechanism of confinement for the gluons 
(see also the discussion after Eq.\ (18b) in \cite{Dokshitzer:2004ie}).

%%%%%%%%%%%%%%%%%%%%%%%%%%%%%%%%%%%%%%%%%%%%%%%%%%%%%%%%%%%%%%%%%%%%%

\section*{ACKNOWLEDGMENTS}

The authors thank Dan Zwanziger for helpful comments.
Research supported by FAPESP (Projects No.\ 00/05047-5 and 03/05259-0).
Partial support (AC, TM) from
% Conselho Nacional de Desenvolvimento Cient\'{\i}fico e Tecnol\'ogico (CNPq)
CNPq is also acknowledged.

%%%%%%%%%%%%%%%%%%%%%%%%%%%%%%%%%%%%%%%%%%%%%%%%%%%%%%%%%%%%%%%%%%%%%%%%%%%%%%%

\end{document}